\documentclass[manuscript]{acmart}
\usepackage{algpseudocode}
\usepackage{multirow}
\usepackage[linesnumbered,ruled,vlined]{algorithm2e}

\begin{document}
\title{Hierarchical Reinforcement Learning for Modeling User Novelty-Seeking Intent in Recommender Systems}

%\author{Anonymous for Review}

\author{Pan Li}
\affiliation{%
  \institution{New York University}
  \streetaddress{44 West 4th Street}
  \city{New York}
  \country{USA}}
\email{pli2@stern.nyu.edu}

\author{Yuyan Wang}
\affiliation{%
  \institution{Google Research}
  \streetaddress{Mountain View}
  \city{California}
  \country{USA}}
\email{yuyanw@google.com}

\author{Ed H. Chi}
\affiliation{%
  \institution{Google Research}
  \streetaddress{Mountain View}
  \city{California}
  \country{USA}}
\email{edchi@google.com}

\author{Minmin Chen}
\affiliation{%
  \institution{Google Research}
  \streetaddress{Mountain View}
  \city{California}
  \country{USA}}
\email{minminc@google.com}

\begin{abstract}
Recommending novel content, which expands user horizons by introducing them to new interests, has been shown to improve users' long-term experience on recommendation platforms \cite{chen2021values}. Users however are not constantly looking to explore novel content. It is therefore crucial to understand their novelty-seeking intent and adjust the recommendation policy accordingly. Most existing literature models a user's propensity to choose novel content or to prefer a more diverse set of recommendations at individual interactions. Hierarchical structure, on the other hand, exists in a user's novelty-seeking intent, which is manifested as a static and intrinsic user preference for seeking novelty along with a dynamic session-based propensity. To this end, we propose a novel hierarchical reinforcement learning-based method to model the hierarchical user novelty-seeking intent, and to adapt the recommendation policy accordingly based on the extracted user novelty-seeking propensity. We further incorporate diversity and novelty-related measurement in the reward function of the hierarchical RL (HRL) agent to encourage user exploration \cite{chen2021values}. We demonstrate the benefits of explicitly modeling hierarchical user novelty-seeking intent in recommendations through extensive experiments on simulated and real-world datasets. In particular, we demonstrate that the effectiveness of our proposed hierarchical RL-based method lies in its ability to capture such hierarchically-structured intent. As a result, the proposed HRL model achieves superior performance on several public datasets, compared with state-of-art baselines.  

\end{abstract}


\maketitle

\section{Introduction}
Recommender system constitutes one of the most important information filtering systems that provide users with the most relevant content. Classic recommendation models focus primarily on matching users with the most relevant items based on their historical activities \cite{adomavicius2005toward}. Recent literature \cite{chen2022off,chen2021values,wang2022surrogate} however pointed out the need for user exploration, when designing recommender systems. In particular, users might get bored with repeated types of item recommendations and would therefore prefer to seek novel content. By producing diversified and unexpected recommendations for the targeted users, we could expand their horizons, address their novelty-seeking desire and improve their online experience as a result. To this end, multiple user exploration-based recommendation models have been proposed, where researchers utilize bandit-based methods \cite{li2010contextual} and reinforcement learning-based methods \cite{afsar2021reinforcement} to exploit existing user interests while simultaneously exploring new user interests. These models have achieved significant recommendation performance improvements, leading to industrial adoption by a number of major recommendation platforms \cite{chen2021values}.

Despite the great success achieved by these user exploration models, understanding the user novelty-seeking intent and adapting recommendation policies accordingly still remains challenging and under-explored. Specifically, existing methods focus on identifying suitable exploratory content for the user, without explicitly modeling the propensity of each user to select novel and diverse content. There is a missed opportunity here as user novelty-seeking intent can be affected by both the static, intrinsic user preference and a dynamic session-based factor. As a result, the user novelty-seeking intent can vary significantly across users, and for the same user, it can fluctuate across different recommendation sessions or even within the same recommendation session. For example, a power user with broad interests could be content with consuming only content from their known interests, while a new user could be drawn to explore and discover new interests. Meanwhile, the same user could enjoy browsing curiosity-inducing content during the morning rush hour on public transportation to kill time, while on a relaxing Saturday night, he or she might prefer to pick up the TV series where they left off last week. In another example, the user might want to switch to other genres, after binge-watching the same TV series. The complex and hierarchical structure of user novelty-seeking intent has not been explicitly and systematically captured in the existing methods.

To this end, we propose a novel hierarchical reinforcement learning (HRL) method to model the user novelty-seeking intent in recommendations, and to update the recommendation policy accordingly. Specifically, motivated by the Deep Reinforcement Learning technique \cite{mnih2015human}, we formulate two modules in our proposed recommendation model: (1) the Session-Level DDPG, which captures high-level, abstract, and session-based user novelty-seeking intent. It stays static throughout the whole session, and gets updated when the user enters a new recommendation session. (2) the Interaction-Level DQN, which captures the dynamic and personalized user novelty-seeking intent towards each item. It is updated upon every new interaction between a user and an item. By taking into account both the session-level and interaction-level novelty-seeking intent when modeling a user's decision-making process, our proposed method is able to produce more effective user exploration strategies and superior recommendation performance.

We in addition study the design of reward functions in our proposed method. We found it beneficial to explicitly incorporate novelty and diversity-based metrics in the reward functions for optimizing the Session-Level DDPG and the Interaction-Level DQN respectively, in order to further encourage user interest exploration. We validate the effectiveness of our proposed method on a simulation dataset and three real-world industrial datasets, where it achieves significant recommendation performance improvement over selected state-of-the-art recommendation baselines. We also conducted an extensive set of ablation studies to understand the importance of each component in our proposed method.

In summary, we make the following research contributions in this paper:
\begin{itemize}
\item We provide empirical evidence to demonstrate the importance of modeling hierarchical user novelty-seeking intent in the design of recommender systems.
\item We propose a novel hierarchical reinforcement learning-based recommendation model that consists of a Session-level DDPG and an Interaction-level DQN to capture the hierarchical user novelty-seeking intent and adapt the recommendation policy accordingly.
\item We test our proposed method through extensive simulation and offline experiments, showcasing that our model can effectively capture hierarchical user novelty-seeking intent in recommendations, and achieves significant performance improvement over state-of-the-art baselines.
\end{itemize}

\section{Related Work}
\subsection{User Novelty-Seeking Intent in Recommendations}
Classic recommendation models focus primarily on matching contents similar to known user interests \cite{adomavicius2005toward,xiao2007commerce}. They often overlook the dispersion of the user’s recommendations, as well as users’ desire to seek novel recommendations \cite{givon1984variety,adamopoulos2014unexpectedness}. As a matter of fact, users can be interested in the unshown items on the platform. Optimizing purely the relevance metric can lead to feedback-loop biases \cite{sun2019debiasing} and the filter bubble phenomenon \cite{pariser2011filter}, reducing user satisfaction in the long run.

The ''exploration-exploitation'' trade-off \cite{balabanovic1998exploring} has been extensively studied in the context of recommender systems. The recommendation agent faces the dilemma of determining whether to exploit the known user interests by recommending items similar to their historical consumption, or to explore new user interests by recommending novel items. Multi-arm bandits \cite{li2010contextual,auer2002finite,chapelle2011empirical} have long been used to make the trade-off, in which one allocates a certain proportion of online traffic for exploring user interests while exploiting the known interests in the rest of the traffic \cite{agrawal2012analysis}. Meanwhile, reinforcement learning techniques \cite{mnih2015human} formulate the Markov Decision Process (MDP) \cite{shani2005mdp} for decision making. These RL-based recommendation models, which aim at identifying optimal recommendation actions to maximize the long-term objectives, such as user retention rates and churn rates \cite{zhao2019deep}, make the trade-off implicitly as well. Some representative deep reinforcement learning-based recommendation models include \cite{zheng2018drn,zou2019reinforcement,xin2020self,chen2021values}.

While these user exploration models have achieved great success in various recommendation applications, they only focus on identifying suitable exploratory content for the user, without properly modeling the propensity of each user to select novel and diverse content. In particular, they do not explicitly capture the hierarchical structure of user novelty-seeking intent. As a result, the long horizon span of multi-session online user interactions can easily render these methods sub-optimal. In this paper, we propose a novel hierarchical reinforcement learning-based model to capture the hierarchical user novelty-seeking intent and achieve significantly better performance as a result.

\subsection{Reinforcement Learning for Recommendation}
As discussed in \cite{mnih2015human,sutton2018reinforcement}, reinforcement learning provides a mathematical framework to capture dynamic user preferences and learn recommendation policy to optimize long-term business objectives \cite{zheng2018drn,zou2019reinforcement}, and many reinforcement learning-based models have been proposed for recommendation purposes \cite{xin2020self,zhang2022multi,antaris2021sequence,xie2021hierarchical,xiao2021general}. For example, \citet{liebman2019right} proposed a reinforcement learning framework to generate playlists according to the current context by adapting to a listener's sequential preferences within a listening session. Our work is significantly different from all existing work in that, we explicitly model hierarchical user novelty-seeking intent through a reinforcement learning-based method to produce more effective exploration strategies in recommendations. Deep reinforcement learning combines the modeling capacity of deep neural networks and the MDP formulation of classic reinforcement learning have achieved great success in many domains \cite{chen2019generative,afsar2021reinforcement}. 

\subsection{Hierarchical Reinforcement Learning}
Hierarchical Reinforcement Learning intends to address the sample inefficiency of RL, especially in long-horizon problems, through temporal abstraction. One class of HRL methods integrates hierarchical action-value functions that operate at different temporal scales \cite{kulkarni2016hierarchical}. A top-level q-value function learns a policy over intrinsic goals, while a lower-level function learns a policy over atomic actions to satisfy the given goals. It allows for flexible goal specifications, such as functions over entities and relations, and provides an efficient space for exploration in complicated environments. It has achieved great success across different applications, such as robotics \cite{kulkarni2016hierarchical}, arcade learning \cite{bacon2017option}, self-driving \cite{codevilla2018end}, and data-efficient learning \cite{nachum2018data}. Motivated by the effectiveness of hierarchical reinforcement learning methods, we propose a novel recommendation model to capture the hierarchical user novelty-seeking intent by estimating the Q-value in each recommendation session.

\section{Method}
\begin{figure*}
\centering
\includegraphics[width=\textwidth]{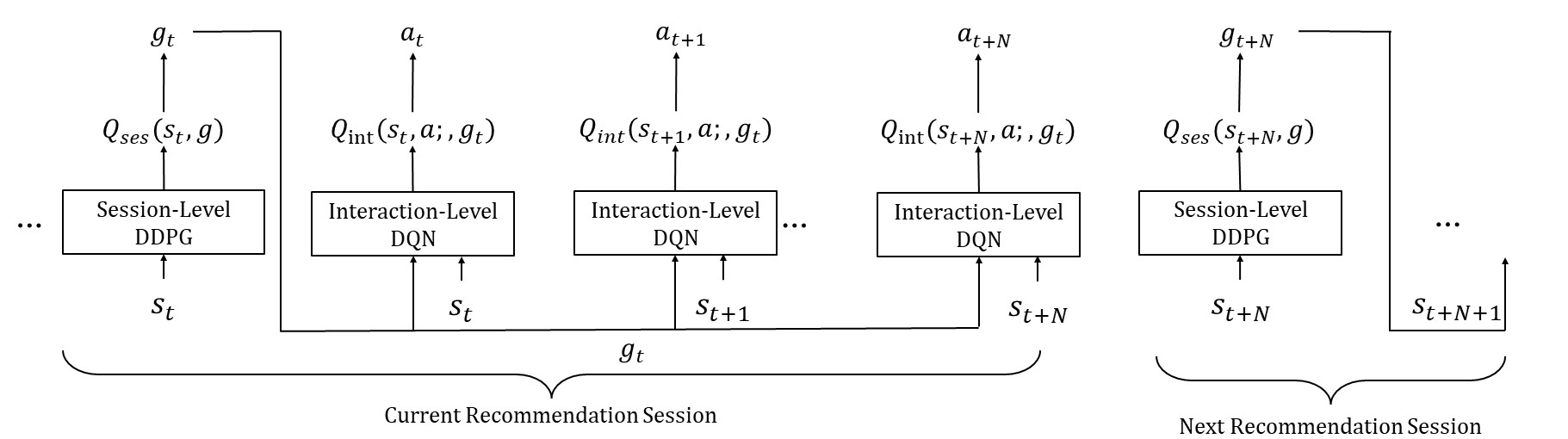}
\caption{Overview of the proposed Hierarchical Reinforcement Learning-Based Model.}
\label{model}
\end{figure*}

In this section, we introduce the hierarchical reinforcement learning-based model to capture the hierarchical user novelty-seeking intent, and to improve recommendation quality. We first formulate the problem under the Q-Learning framework, and present the Hierarchical Q-learning solution. We then describe the learning process of the Session-DDPG and Interaction-DQN respectively. Finally, we discuss the design of reward functions and summarize our proposed model. An overview of the model is shown in Figure \ref{model}.

\subsection{The Hierarchical Reinforcement Learning Framework}
To start with, we formulate the task of determining the optimal recommendation policy as a Markov Decision Process (MDP), and use Q-Learning \cite{watkins1992q} to learn the policy. The Q-value of taking action $a$ in state $s$ under a recommendation policy $\pi$ equals to
\begin{equation}
Q^\pi(s_t;a_t) = E_{\tau\sim p_\pi(\tau|s_t, a_t) }[\sum_{t'=t} \gamma^{t'-t} r(s_{t'}, a_{t'})]
\end{equation}
where $r(s_t, a_t)$ represents the immediate reward for taking action $a_t$ under state $s_t$, and the discount factor $\gamma$ controls the relative importance of the immediate reward and the future reward. At each step $t$, the agent chooses the action that maximizes the Q-value, i.e., $a^\ast_t=\arg\max_a{Q^\pi(s_t;a_t)}$. 

To estimate the Q-value in each recommendation session, and to capture the hierarchical user novelty-seeking intent in recommendations, we adopt the Deep Hierarchical Reinforcement Learning method from the robotics literature \cite{kulkarni2016hierarchical}, and apply it to the recommendation tasks in a novel manner. The tabular setting in classic reinforcement learning cannot handle the enormous state and action spaces in recommendation settings. Furthermore, the highly complex and nonlinear nature of the values of user-item interactions requires high-capacity models. The deep neural network, on the other hand, is capable of learning personalized and dynamic user-item relationships in recommendations. In this work, we follow the Deep Q-Network (DQN) technique \cite{mnih2015human} and the Deep Deterministic Policy Gradient (DDPG) technique \cite{lillicrap2015continuous} that both utilize deep neural networks to estimate the Q-function. 

Specifically, our model consists of the following two modules: (1) the Session-Level DDPG, which captures the high-level, abstract, and personalized user novelty-seeking intent in each session; and (2) the Interaction-Level DQN, which captures the dynamic and personalized user novelty-seeking intent at each item interaction. The Session-Level DDPG takes the initial user state at the beginning of each session as the input, and produces the session policy $g_t$, which is a latent vector that controls the overall recommendation policy within the current recommendation session. The Session-Level DDPG and the latent policy vector $g_t$ will stay static through the same session, and only get updated when the user enters a new recommendation session. We explain the motivation of the design choice in Section 
\ref{sec:learning_process} below. The Interaction-Level DQN takes the dynamic user state as the input, and produces the optimal action $a_t$ as the recommended product that will be provided to the consumer. As the result, the Interaction-Level DQN and the produced action $a_t$ will be updated upon every user-item interaction. By considering both the session-level and interaction-level user novelty-seeking intent when modeling users' decision-making process, our algorithm is able to produce more effective user exploration strategies and better recommendations. Our proposed deep hierarchical reinforcement learning-based model is visualized in Figure \ref{model}.

We use a Dueling Double Deep Q-Network, which combines two variants of the DQN, namely Double DQN \cite{van2016deep} and Dueling DQN \cite{wang2016dueling} to construct both the Session-Level DDPG and the Interaction-Level DQNs. The Dueling DQN decomposes the Q-value estimation into two separate components: $V(s)$, the value of being in state $s$; and $A(s;a)$, the advantage of taking action $a$ in state $s$. This decomposition leads to better learning efficiency, more accurate Q estimation, and better policy than the traditional DQN, as discussed in \cite{wang2016dueling}. The Double DQN introduces a separate target network $Q'$ for target Q-value generation apart from the $Q$ network used for action selection. The $Q$ network is the primary network that is updated at every learning step, while the target network $Q'$ adopts periodical syncing to the $Q$ network. Decoupling the two and allowing the slow update to the target network reduces the overestimation of Q-values and stabilizes learning.

\subsection{The Learning Process}
\label{sec:learning_process}
As shown in Figure~\ref{model}, the Session-level DDPG learns a session policy or goal $g_t$ that supervises the interaction-level policy within this session. Note that $g_t$ is a continuous, latent vector, and we cannot traverse all possible options of $g_t$ to select the optimal values, as was done in the classical DQN settings. Therefore, we follow the learning paradigm in DDPG \cite{lillicrap2015continuous} and produce the optimal session policy $g_t$ through a separate action network $\mu_\phi$. The action network consists of multiple fully-connected hidden layers, which take the user state as the input and output a latent vector $g_t$ as the goal. The parameters of the action network will be updated through back-propagation when we minimize the Temporal Difference (TD) loss described below.

When a user arrived at the beginning of a recommendation session $t$ with state $s_t$, We learn the Session-level DDPG to maximize the expected session return, which can be expressed by the following Bellman equation:
\begin{equation}
\begin{split}
Q_{\theta_{ses}}(s_t, g_t)=r_{ses}(s_t, g_t) +\gamma Q'_{\theta'_{ses}}(s_{t+N}, g_{t+N}) \\
\mbox{ where } g_t = \mu_{\phi_{ses}}(s_t), \ g_{t+N} = \mu_{\phi'_{ses}}(s_{t+N})
\end{split}
\label{eq:q_session}
\end{equation}
where $g_t \in \mathcal{R}^d$ is a latent vector representing session-level policy and will be learned through back-propagation from the action network $\mu_{\phi_{ses}}$; $g_{t+N}$ is the session policy vector for the next recommendation session, which is learned from the target action network  $\mu_{\phi'_{ses}}$ by polyak averaging the action network parameters over the course of training; $r_{ses}(s_t, g_t)$ represents the session-level immediate reward for taking session goal $g_t$; $s_{t+N}$ is the user state representation at the beginning of next recommendation session, when acting according to $g_t$ in the current session; $Q_{\theta_{ses}}$ and $Q'_{\theta'_{ses}}$ are the primary and target networks in the Session-Level DDQG model that we have previously described. 

Following the Double Dueling DQN design, we further decompose $Q_{\theta_{ses}}(s_t, a_t)$ as
\begin{equation}
\begin{split}
Q_{\theta_{ses}}(s_t, g_t)=V_{\theta_{ses}}(s_t) + A_{\theta_{ses}}(s_t, g_t)
\end{split}
\label{eq:q_session_double}
\end{equation}
Temporal Different (TD) learning is employed to learn the parameters of the Q networks and action networks. 
\begin{equation}
\begin{split}
&\ell_{ses}(\theta_{ses}, \phi_{ses}, \theta'_{ses}, \phi'_{ses}) = \\
&\left[Q_{\theta_{ses}}(s_t, g_t) - \left(r_{ses}(s_t, g_t) +\gamma Q'_{\theta'_{ses}}(s_{t+N}, g_{t+N})\right)\right]^2
\end{split}
\label{eq:l_session}
\end{equation}
We further plug-in equation~(\ref{eq:q_session_double}) to replace the Q networks with the value networks (V networks) and advantage networks (A networks), and optimize these two networks accordingly to produce the Q-value estimation for each action $g_t$ under user state $s_t$.

Denote $\mathcal{A}$ as the discrete action space and $a\in \mathcal{A}$ as the recommended item. We pick $a_t$ according to the interaction-level DQN, conditioning on the Session-level policy $g_t$. By conditioning, we concatenate the latent vector $g_t$ to the state vector $s_t$. The interaction-level Q value, similarly, follows the Bellman equation as follows:
\begin{equation}
\begin{split}
Q_{\theta_{int}}(s_t, a_t; g_t)  =   r_{int}(s_t, a_t) + \gamma Q'_{\theta'_{int}}(s_{t+1}, a_{t+1}^\ast; g_t)\\
\mbox{ where } a_{t+1}^\ast = \arg\max_{a'} Q'_{\theta'_{int}}(s_{t+1}, a';g_t) 
\end{split}
\end{equation}
Here $r_{int}(s_t, a_t)$ represents the interaction-level immediate reward. $s_{t+1}$ is the user state transitioned from $s_t$ when taking action $a_t$. We again use dueling to decompose the $Q_{\theta_{int}}(s_t, a_t; g_t)$, and learn the interaction-level value network $V_{\theta_{int}}(s_t)$ and advantage network $A_{\theta_{int}}(s_t, a_t)$ using TD learning.
\begin{equation}
\begin{split}
&\ell_{int}(\theta_{int}, \theta'_{int}) =\\ &\left[Q_{\theta_{int}}(s_t, a_t; g_t)  - \left( r_{int}(s_t, a_t) + \gamma Q'_{\theta'_{int}}(s_{t+1}, a_{t+1}^\ast; g_t)\right)\right]^2
\end{split}
\end{equation}

We learn the Session-level DDPG and the Interaction-level DQN by jointly minimizing the combined TD losses. 
\begin{equation}
\begin{split}
\ell = \ell_{ses}(\theta_{ses}, \phi_{ses}, \theta'_{ses}, \phi'_{ses}) + \ell_{int}(\theta_{int}, \theta'_{int})
\end{split}
\end{equation}
The user states are encoded through the GRU network \cite{chung2014empirical}, where we feed the explicit user features and user-item interaction history as inputs, and produce the dynamic user state representations from the sequential neural network accordingly. 

\subsection{The Reward Function}
To enable learning of our proposed model, we need to formulate the session-level reward $r_{ses}(s_t, g_t)$ and the interaction-level reward $r_{int}(s_t, a_t)$ based on the user-item interaction records. As user reward cannot be explicitly observed in the offline experiment settings, existing literature \cite{afsar2021reinforcement} typically use the average item ratings in a recommendation session as the proxy of the session-level reward $r_{ses}(s_t, g_t)$, and the item rating as the proxy of the interaction-level reward $r_{int}(s_t, a_t)$. However, these reward designs do not capture the exploration-related objectives of the user. To further encourage user exploration in recommendations, we propose to incorporate novelty-based and diversity-based metrics in the reward functions of our proposed deep hierarchical reinforcement learning-based model. In particular, the session-level reward $r_{ses}(s_t, g_t)$ is formulated as
\begin{equation}
\label{eqn:ses_r}
r_{ses}(s_t, g_t) = R_{ses}(s_t, g_t) + D_{ses}(s_t, g_t),
\end{equation}
and the interaction-level reward $r_{int}(s_t, a_t)$ is formulated as 
\begin{equation}
\label{eqn:int_r}
r_{int}(s_t, a_t) = R_{int}(s_t, a_t)+N_{int}(s_t, a_t),
\end{equation}
where $R_{ses}$ is the average rating of all the interactions within a session, and $R_{int}$ is the rating of the item in the interaction. $D_{ses}$ computes the average pairwise dissimilarity\footnote{See Section \ref{sec:simulation_settings} below for the dissimilarity measures.} within the list of items in the current recommendation session, and $N_{int}$ represents the deviation of the current recommended item from the last item consumed by the user\footnote{See Section \ref{sec:simulation_settings} below for the novelty measures.}. We empirically validated these designs through extensive experiments on the simulation datasets and three real-world datasets. In the experiments, we will show that explicitly incorporating novelty-based and diversity-based metrics in the reward functions significantly improves user exploration. 

\section{Experiment on Simulation Datasets}
To demonstrate the effectiveness of our proposed model, we conduct a series of experiments on the simulation dataset, and compare its performance with selected state-of-the-art recommendation baselines. We also conducted extensive ablation experiments to shed light on the importance of different components. We will now start by introducing the setup for our simulation experiment.

\subsection{Simulation Experiment Settings}
\label{sec:simulation_settings}
In our simulation experiment, the user states will be dynamically updated as the user interacts with the recommendation agent. Specifically, at \emph{each} timestamp $t$, we generate the interaction-level reward from the selected item $i$ for user $u$ using the discrete choice model in the item response theory \cite{embretson2013item} as: 
\begin{equation}
\label{eqn:R}
r_{ui,t} = A_{ui} + E_{u,t} * N_{ui} + r_{u} + r_{i} + e_{ui,t}
\end{equation}
where $r_{ui,t}$ represents the simulated reward of recommending item $i$ to user $u$ at time $t$, $A_{ui}$ stands for the relevance (affinity) score between user $u$ and item $i$, $E_{u,t}$ is a continuous value capturing the novelty-seeking intent of the user in the current recommendation session, of which the generation process is explained below. $N_{ui}$ is the novelty score of item $i$ with regard to the user $u$, $r_{u}$ and $r_{i}$ represent the intrinsic rating levels (fixed effects) for user $u$ and item $i$, respectively. $e_{ui,t}$ is the random bias term that is added to the model to simulate the fluctuation of product utility values for each consumer in practice. In our simulation experiment, these variables are all sampled from normal distributions with predetermined mean and variance values. Specifically, we draw $r_{u} \backsim N(0.5,1)$, $r_{i} \backsim N(0.5,1)$ and $e_{uit} \backsim N(0,0.1)$. 

Meanwhile, the relevance objective in our simulation experiment is modeled as 
\begin{equation}
\label{eqn:A}
% A_{ui} = 1 - ||e_{u} - e_{i} ||^2,
A_{ui} = e_{u} ^T e_{i},
\end{equation}
where $e_u$ and $e_i$ constitute the latent representations of explicit user features and item features which are normalized to the unit sphere (i.e.$||e_u||_2 = 1$, $||e_i||_2 = 1$). Each dimension of $e_u$ and $e_i$ is sampled from the uniform distribution $U[0,1]$, and the similarity function is defined as the inner product between the user representation and item representation. We also model the user novelty-seeking intent as a combination of user-based intrinsic intent and session-based intent, i.e.,
\begin{equation}
\label{eqn:E}
E_{ut} = E_u^0 + E_t^{s(t)},
\end{equation}
where $E_u^0$ represents the intrinsic user propensity to explore novel items, while $E_t^{s(t)}$ represents the session-level user novelty-seeking intent for the current session $s(t)$. $E_u^0$ is sampled from $N(0,1)$, and $E_t^{s(t)}$ for each session $s(t)$ is also sampled from $N(0,1)$, where the session length is fixed at 5 in our experiments~\footnote{This can be easily generalized to variable session length, or other definitions of session}. We have also conducted additional simulation experiments using session lengths of 10 and 20, and observed similar levels of performance improvements that we illustrate in this paper. Finally, we model the novelty objective as
\begin{equation}
\label{eqn:N}
N_{ui} = ||e_{i} - e_{i_{t-1}}||_2,
\end{equation}
where $e_{i_{t-1}}$ is the embedding of the last consumed item, and $||\cdot||_2$ is the Euclidean distance (item embeddings are normalized to the unit sphere). Intuitively, $N_{ui}$ captures the dissimilarity between the current item and the last consumed item by the user.

In our simulation experiment, we first generate the rating matrix $R_{u,i}$ with 10,000 consumers and 10,000 products, thus having 100,000,000 ratings in total. At step 1, we recommend the top-1 product with the highest reward for each user to be used as the interaction history; and then from step 2 to 40, we construct the training set by recommending one product to the user each time based on the policy from each individual agent; we then simulate the reward as defined in equation (\ref{eqn:R}) assuming the user would interact the recommendation. We summarize the details for generating the trajectories for each user in Algorithm \ref{alg1}, which is an online learning procedure when the agent is learned through our proposed method. Algorithm \ref{alg2} describes step 41-50, where we roll out the learned policy from Algorithm \ref{alg1} for another 10 steps, and compute evaluation metrics as introduced in Section \ref{sec:metrics} below. The evaluation results are reported as the average of 10 independent runs.

%%%%%%%%%%%%%%%%%%%%%%%%%Algo 1%%%%%%%%%%%%%%%%%%%%%%%%%%%%%
\begin{algorithm}[ht]
\SetAlgoNoLine
\DontPrintSemicolon
\KwIn{$e_u$, $e_i$ for all users $u=1,...,U$ and all items $i=1,...,I$; Model parameters $\theta_{ses}$, $\theta'_{ses}$, $\phi_{ses}$, $\phi'_{ses}$ for the session-level DDPG network, and $\theta_{int}$, $\theta'_{int}$ for the interaction-level DQN; Reward discounting factor $\gamma$; Learning rate $\alpha$; Session length $L=5$; Training steps $n_{train}$ = 40.} 
 Initiate the networks $Q_{\theta_{ses}}$ and $Q'_{\theta_{ses}}$ for session-level DDPG, $Q_{\theta_{int}}$ and $Q'_{\theta_{int}}$ for interaction-level DQN, and initial state $s_{u1}$, $g_1$. \;
 \For{$u=1,...,U$}{
     \For{$t=1,...,n_{train}$}{
     Obtain $g_t$ from the action network $\mu_\phi(s_{ut})$ for $Q_{\theta_{ses}}$, where $s_{ut}$ is the user state representation obtained through the GRU network\;
     Pick action $a_t = \text{argmax}_{a'}Q_{\theta_{int}}(s_{ut}, a'; g_t)$. \;
     Observe $r^{int}_{ui,t}$ according to Eq.(\ref{eqn:int_r}). \;
     Update the Interaction-level DQN parameters $\theta_{int}$ through TD learning using the newly observed sample $(s_{u,t-1}, a_{t-1}, r^{int}_{ui,t}, a_t)$, and the target network parameters $\theta'_{int}$ using polyak averaging. \;
     \If{$t \% L = 0$}{
     Compute session-level reward $r^{ses}_{uit}$ as in Eq.(\ref{eqn:ses_r}). \;
     Update the session-level DDPG parameters $\theta_{ses}, \phi_{ses}$ through TD learning using the newly observed sample $(s_{u,t-L}, g_{t-L}, r^{ses}_{ui,t}, g_{t})$, and the target network parameters $\theta'_{ses},  \phi'_{ses}$ using polyak averaging. \;
     }
     } 
 }
\KwOut{Updated model parameters $\theta_{ses}$, $\theta'_{ses}$, $\phi_{ses}$, $\phi'_{ses}$,  $\theta_{int}$ and $\theta'_{int}$.}
\caption{Simulating the Training Data}
\label{alg1}
\end{algorithm}

%%%%%%%%%%%%%%%%%%%%%%%%%Algo 2%%%%%%%%%%%%%%%%%%%%%%%%%%%%%
\begin{algorithm}[ht]
\SetAlgoNoLine
\DontPrintSemicolon

\KwIn{Learned parameters $\theta_{ses}$, $\theta'_{ses}$, $\phi_{ses}$, $\phi'_{ses}$,  $\theta_{int}$ and $\theta'_{int}$; $e_u$, $e_i$ for all users $u=1,...,U$ and all items $i=1,...,I$; K=10 as number of recommendations per user; Number of test steps $n_{test} = 10$.}  
\For{$t=n_{train} + 1,...,n_{train} + n_{test}$}{ 
  Get $s_t$ from the GRU component and Obtain $g_t$ from session-level action network $\mu_{\phi_{ses}}(s_t)$, and  therein. \;
  Generate the predicted Q-values for every action $a_i$, i.e. $Q_{\theta_{int}}(s_t, a_t; g_t)$. \;
  Select the top K actions $a^*_1,...,a^*_{K}$ based on the Q-values. \;
  Sample $r_u$ $\backsim N(0.5,1)$\;
     \For{$i\in \{a^*_1,...,a^*_{K}$\}}{
     Sample $r_i \backsim N(0.5,1)$; Compute $A_{ui}$, $E_{u,t}$ and  $N_{ui}$ according to Eq (\ref{eqn:A}), Eq (\ref{eqn:E}) and Eq (\ref{eqn:N}); Sample $e_{ui,t} \backsim N(0,0.1)$. \;
     } 
  Obtain ground-truth rewards for the top K actions based on Eq.(\ref{eqn:R})\;
  Compute Average Reward, Hit Rate@10, Diversity, and Novelty metrics defined in Section \ref{sec:metrics}. \;
  Add the top-1 item into the history of the user, and update the user state representation $s_t$ through the GRU network accordingly.
}
 
\KwOut{Average Reward, Hit Rate@10, Diversity and Novelty metrics.}
\caption{Simulating the Test Data (for a single user).}
\label{alg2}
\end{algorithm}

\subsection{Simulation Experiment Baselines and Metrics}
\label{sec:metrics}
To demonstrate the effectiveness of our proposed model, we compare its performance with selected state-of-the-art recommendation baselines, ranging from relevance-oriented baseline methods of DIN, DeepFM, Wide \& Deep, and PNN, to state-of-the-art reinforcement learning-based baseline methods of HRL-Rec, REINFORCE and DRN. We summarize these baseline models below:
\begin{itemize}
\item \textbf{HRL-Rec \cite{xie2021hierarchical}} The Hierarchical reinforcement learning framework for integrated recommendation (HRL-Rec) model produces the integrated recommendation into two agents: the low-level agent is a channel selector, which generates a personalized channel list; while the high-level agent is an item recommender, which recommends specific items from heterogeneous channels under the channel constraints.
\item \textbf{REINFORCE \cite{sutton2018reinforcement}} The classic REINFORCE algorithm, which has been successfully applied in a large-scale commercial recommender system~\cite{chen2019top} with the additional off-policy correction to address data biases in learning from logged feedback.
\item \textbf{DRN \cite{zheng2018drn}} The DRN model is constructed based on Deep Q-Learning, which explicitly models future rewards. The DRN model also considers the user return pattern as a supplement to clicking labels to capture more user feedback information and to establish an effective exploration strategy.
\item \textbf{DIN \cite{zhou2018deep}} Deep Interest Network designs a local activation unit to adaptively learn the representation of user interests from historical behaviors with respect to a certain item.
\item \textbf{DeepFM \cite{guo2017deepfm}} DeepFM combines the power of factorization machines for recommendation and deep learning for feature learning in a new neural network architecture.
\item \textbf{Wide \& Deep \cite{cheng2016wide}} Wide \& Deep utilizes the wide model to handle the manually designed cross-product features, and the deep model to extract nonlinear relations among features.
\item \textbf{PNN \cite{qu2016product}} Product-based Neural Network model introduces an additional product layer to serve as the feature extractor.
\end{itemize}

In addition, we compare our method with several bandit-based recommendation models that could balance between exploration and exploitation strategies in recommendations, which include the following:

\begin{itemize}
\item \textbf{LinUCB \cite{li2010contextual}} LinUCB models the personalized recommendation task as a contextual bandit problem, where the learning algorithm sequentially selects items to serve users based on contextual information, while simultaneously adapting its strategy based on user feedback to maximize total rewards.
\item \textbf{TS \cite{chapelle2011empirical}} Thompson Sampling (TS) is a method for choosing actions to address exploration-exploitation in the multi-armed bandit problem by choosing the action that maximizes the expected reward with respect to a randomly drawn belief.
\item \textbf{COFIBA \cite{li2016collaborative}} The Collaborative Filtering Bandit method takes into account the collaborative effects that arise due to the interaction of the users with the items, and also takes advantage of preference patterns in the data in the bandit-learning process.
\end{itemize}

Finally, we also construct multiple variants of our proposed model to shed light on the importance of each component in our model design:
\begin{itemize}
\item \textbf{Ablation 1 (No Session Intent)} In this ablation model, the user novelty-seeking intent is solely determined by the user's intrinsic novelty-seeking level, without taking into account the session-based novelty-seeking intent.
\item \textbf{Ablation 2 (No Hierarchical RL)} In this ablation model, the recommendations are provided following the classical reinforcement learning method, instead of the hierarchical reinforcement learning method. That is to say, we remove the Session-DDPG from our proposed model.
\item \textbf{Ablation 3 (No Hierarchical RL+No Session Intent)} In this ablation model, we remove the Session-DDPG from our proposed model and formulate the user novelty-seeking intent only based on the user's intrinsic exploration level.
\item \textbf{Ablation 4 (Vanilla DQN)} In this ablation model, we formulate the networks of both Session-DDPG and Interaction-DQN through the vanilla DQN method, instead of the Double Dueling DQN method in our proposed model.
\item \textbf{Ablation 5 (No Novelty)} In this ablation model, we remove the novelty metric from the interaction-level reward, i.e. $r_{int}(s_t, a_t) = R_{int}(s_t, a_t)$ in Eq.(\ref{eqn:int_r}).
\item \textbf{Ablation 6 (No Diversity)} In this ablation model, we remove the diversity metric from the session-level reward, i.e. $r_{ses}(s_t, g_t) = R_{ses}(s_t, g_t)$ in Eq.(\ref{eqn:ses_r}).
\end{itemize}

To evaluate the recommendation performance in the simulation experiment, we consider the following four metrics:
\begin{itemize}
\item \textbf{Average Reward}, which measures the average reward that the user could get from the top-K (K=10) item recommendations generated by the recommender system. The reward values are simulated following Eq.(\ref{eqn:R}).
\item \textbf{Hit Rate@K}, which measures the percentage of ''positive'' items (reward greater than 0.5) in the top-K item recommendations generated by the recommender system. 
\item \textbf{Diversity}, which measures the pairwise dissimilarity in the top-K item recommendations, calculated as the Euclidean distance between their latent embeddings.
\item \textbf{Novelty}, which measures the deviation of the current item recommendation to the last purchased item of the user, calculated as 1 minus the Euclidean distance between their latent embeddings.
\end{itemize}

\subsection{Simulation Experiment Results}
\label{sec:simulation_results}
We present the results of the simulation experiment in Table \ref{simulation}. We see that our proposed model has achieved significantly better performance over all selected state-of-the-art recommendation baselines in terms of all four evaluation metrics (Average Reward, Hit Rate@10, Diversity, Novelty). These superior recommendation performance results indicate that it is beneficial to explicitly model the hierarchical user novelty-seeking intent when producing recommendations, and that our proposed model could effectively capture such novelty-seeking intent to provide significantly better recommendation performance. We also verify that there is no significant difference in terms of computational resources between our proposed model and the baseline methods.

In addition, our proposed model also achieves significant and consistent performance improvements over the multiple ablation variants. By comparing the recommendation performance of our model with Ablation 1, we verify that the session-level novelty-seeking intent is an important component that needs to be properly modeled in the recommendation process. Ablation studies 2-3 further confirm that hierarchical reinforcement learning is an essential component in our proposed model, removing which resulted in significantly worsened recommendation performance. Compared with Ablation 4, we can see introducing Double Dueling DQNs to reduce over-estimation and improve stability in the classic DQN leads to better recommendation policy. Finally, ablation settings 5 and 6 verify the hypothesis in the paper that it is important to explicitly incorporate the novelty-based metric and diversity-based metric in reward functions.

\begin{table}
\centering
\resizebox{0.5\textwidth}{!}{
\begin{tabular}{|c|cccc|} \hline
Algorithm & Avg-Reward & HR@10 & Diversity & Novelty \\ \hline
\textbf{Our Model} & \textbf{0.5080*} & \textbf{0.5270*} & \textbf{0.2567*} & \textbf{0.2600*} \\ 
 & (0.0007) & (0.0010) & (0.0005) & (0.0005) \\
(Improvement over the best below) & +0.36\% & +1.25\% & +2.56\% & +2.85\% \\ \hline
HRL-Rec & 0.4916 & 0.4930 & 0.2496 & 0.2510 \\
REINFORCE & 0.4924 & 0.4948 & 0.2501 & 0.2526 \\
DRN & 0.4877 & 0.4774 & 0.2488 & 0.2496 \\
DIN & 0.5012 & 0.5132 & 0.1734 & 0.1895 \\
DeepFM & 0.5003 & 0.5118 & 0.1730 & 0.1906 \\
Wide \& Deep & 0.4916 & 0.5096 & 0.1728 & 0.1925 \\
PNN & 0.4905 & 0.5077 & 0.1736 & 0.1925 \\ \hline
LinUCB & 0.4773 & 0.4882 & 0.2488 & 0.2489 \\
TS & 0.4816 & 0.4916 & 0.2462 & 0.2501 \\
COFIBA & 0.4879 & 0.4997 & 0.2506 & 0.2473 \\ \hline
Ablation 1 & 0.5044 & 0.5169 & \underline{0.2503} & \underline{0.2528} \\
Ablation 2 & \underline{0.5062} & \underline{0.5205} & 0.2460 & 0.2509 \\
Ablation 3 & 0.5034 & 0.5170 & 0.2472 & 0.2506 \\ 
Ablation 4 & 0.4978 & 0.4147 & 0.2496 & 0.2496 \\ 
Ablation 5 & 0.4998 & 0.5098 & 0.2275 & 0.2428 \\
Ablation 6 & 0.4972 & 0.5066 & 0.2314 & 0.2444 \\ \hline
\end{tabular}
}
\newline
\caption{Comparison of recommendation performance in the simulation dataset. `*'  represents statistical significance at the 0.95 level. Improvement percentages are computed over the best baseline model (including the ablation studies) for each metric.}
\label{simulation}
\end{table}

\subsection{Visualization}
\begin{figure}
\centering
\includegraphics[width=0.5\textwidth]{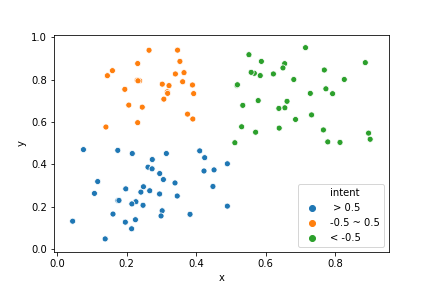}
\caption{TSNE Visualization of the Session Policy}
\label{tsne}
\end{figure}

To further demonstrate that our proposed hierarchical reinforcement learning-based method indeed captures the intended construct of hierarchical user novelty-seeking intent, we visualize the session policy $g_{t}$ of 100 randomly selected users from the simulation experiment (Figure \ref{tsne}). The visualization is achieved using the TSNE technique \cite{van2008visualizing}, which maps the high-dimensional latent session policy vector $g_{t}$ into the 2-dimensional latent space, in such a way that similar objects are modeled by nearby points and dissimilar objects are modeled by distant points with high probability. We then compare the session policy $g_{t}$ with the ground truth user novelty-seeking intent in each recommendation session in our simulation dataset. For better visualization purposes, we classify the ground truth intent into three categories based on the intent values: "novelty-seeking intent > 0.5", "novelty-seeking intent < -0.5", and "-0.5 < novelty-seeking intent < 0.5". As we show in Figure \ref{tsne}, the TSNE visualization of the session policy matches well with the hierarchical user novelty-seeking intent. In particular, the sessions with high, medium, and low novelty-seeking intent are clearly clustered into three disjoint groups. This means that the learned session-level policy $g_{t}$ is indeed capturing the session-level novelty-seeking intent. This validates the effectiveness of our proposed hierarchical reinforcement learning model in capturing the hierarchical structures of user novelty-seeking intent.

\section{Experiment on Real-World Datasets}
\subsection{Data and Experiment Settings}
We further test our model on three real-world datasets: the Yelp Challenge Dataset \footnote{https://www.yelp.com/dataset/challenge}, which is the Round 8 restaurant review dataset, which contains check-in information of users and restaurants, and the user rating information; the MovieLens Dataset \footnote{https://grouplens.org/datasets/movielens/}, which contains information of the users, movies, and ratings; and the Youku dataset collected from the major online video platform Youku, which contains rich information of users, videos, clicks, and their corresponding features. We list the descriptive statistics of these datasets in Table \ref{statisticalnumber}. We normalize the ratings in Yelp and MovieLens datasets into the scale of between 0 and 1 to construct the reward of recommending each item to the user. 
%\footnote{The model and code have been made publicly available at \url{https://anonymous.4open.science/r/HRL-766B/}}

\begin{table}[h]
\centering
\begin{tabular}{|c|c|c|c|}
\hline
Dataset & \textbf{Yelp} & \textbf{MovieLens} & \textbf{Youku}\\ \hline
\# of Ratings & 2,254,589 & 19,961,113 & 1,806,157 \\ \hline
\# of Users & 76,564 & 138,493 & 46,143 \\ \hline
\# of Items & 75,231 & 15,079 & 53,657 \\ \hline
Sparsity & 0.039\% & 0.956\% & 0.073\% \\ \hline
\end{tabular}
\newline
\caption{Descriptive Statistics of Three Datasets}
\label{statisticalnumber}
\end{table}

\begin{table*}[h!]
\centering
\resizebox{\textwidth}{!}{
\begin{tabular}{|c|cccc|cccc|cccc|} \hline
\multirow{2}{*}{Algorithm} & \multicolumn{4}{c|}{Alibaba-Youku} & \multicolumn{4}{c|}{Yelp} & \multicolumn{4}{c|}{MovieLens} \\ \cline{2-13}
&  Avg-Reward & HR@10 & Diversity & Novelty & Avg-Reward & HR@10 & Diversity & Novelty & Avg-Reward & HR@10 & Diversity & Novelty \\ \hline
\textbf{Our Model} & \textbf{0.4796*} & \textbf{0.4383*} & \textbf{0.2571*} & \textbf{0.2488*} & \textbf{0.4642*} & \textbf{0.3854*} & \textbf{0.2568*} & \textbf{0.2486*} & \textbf{0.4700*} & \textbf{0.4030*} & \textbf{0.2562*} & \textbf{0.2487*} \\ 
 & (0.0009) & (0.0010) & (0.0005) & (0.0005) & (0.0008) & (0.0009) & (0.0005) & (0.0005) & (0.0008) & (0.0009) & (0.0005) & (0.0005) \\
(Improve \%) & +1.89\% & +1.55\% & +5.02\% & +5.42\% & +2.16\% & +2.47\% & +4.86\% & +4.50\% & +2.55\% & +3.71\% & +9.02\% & +2.09\% \\ \hline
HRL-Rec    & 0.4451 & 0.4181 & 0.2443 & 0.2349 & 0.4418 & 0.3639 & 0.2303 & 0.2343 & 0.4428 & 0.3811 & 0.2279 & 0.2306 \\
REINFORCE  & 0.4476 & 0.4189 & \underline{0.2448} & \underline{0.2360} & 0.4429 & 0.3651 & 0.2315 & 0.2364 & 0.4446 & 0.3832 & 0.2305 & 0.2333 \\
DRN        & 0.4438 & 0.4166 & 0.1774 & 0.1752 & 0.4401 & 0.3628 & 0.2306 & 0.2325 & 0.4423 & 0.3810 & 0.2298 & 0.2317 \\
DIN        & 0.4656 & 0.4284 & 0.1776 & 0.1758 & 0.4489 & 0.3677 & 0.1814 & 0.1816 & 0.4521 & 0.3864 & 0.1782 & 0.1754 \\
DeepFM     & 0.4648 & 0.4280 & 0.1793 & 0.1744 & 0.4486 & 0.3676 & 0.1830 & 0.1831 & 0.4526 & 0.3857 & 0.1782 & 0.1756 \\
Wide \& Deep & 0.4642 & 0.4271 & 0.1788 & 0.1766 & 0.4473 & 0.3668 & 0.1832 & 0.1830 & 0.4517 & 0.3857 & 0.1784 & 0.1770 \\
PNN        & 0.4610 & 0.4255 & 0.1806 & 0.1795 & 0.4451 & 0.3651 & 0.1855 & 0.1859 & 0.4508 & 0.3854 & 0.1790 & 0.1774 \\ \hline
LinUCB & 0.4374 & 0.4166 & 0.2440 & 0.2338 & 0.4387 & 0.3619 & 0.2301 & 0.2330 & 0.4428 & 0.3832 & 0.2310 & 0.2321 \\
TS & 0.4396 & 0.4169 & 0.2440 & 0.2316 & 0.4395 & 0.3628 & 0.2297 & 0.2341 & 0.4436 & 0.3830 & 0.2316 & 0.2327 \\
COFIBA & 0.4428 & 0.4177 & 0.2412 & 0.2338 & 0.4417 & 0.3628 & 0.2301 & 0.2347 & 0.4436 & 0.3841 & 0.2316 & 0.2321 \\ \hline
Ablation 1 & 0.4688 & 0.4287 & 0.2430 & 0.2336 & 0.4521 & 0.3742 & 0.2428 & 0.2371 & 0.4561 & 0.3816 & 0.2338 & 0.2430 \\
Ablation 2 & \underline{0.4707} & \underline{0.4316} & 0.2445 & 0.2358 & \underline{0.4544} & \underline{0.3761} & \underline{0.2449} & \underline{0.2379} & \underline{0.4583} & \underline{0.3886} & 0.2349 & 0.2435 \\
Ablation 3 & 0.4670 & 0.4270 & 0.2426 & 0.2341 & 0.4508 & 0.3740 & 0.2420 & 0.2376 & 0.4540 & 0.3878 & \underline{0.2350} & \underline{0.2436} \\ 
Ablation 4 & 0.4643 & 0.4301 & 0.2306 & 0.2289 & 0.4516 & 0.3738 & 0.2299 & 0.2287 & 0.4544 & 0.3833 & 0.2315 & 0.2418 \\
Ablation 5 & 0.4626 & 0.4289 & 0.1976 & 0.1968 & 0.4484 & 0.3710 & 0.2018 & 0.2016 & 0.4510 & 0.3810 & 0.2074 & 0.2088 \\
Ablation 6 & 0.4608 & 0.4286 & 0.2058 & 0.2012 & 0.4450 & 0.3701 & 0.2034 & 0.2041 & 0.4499 & 0.3796 & 0.2063 & 0.2071 \\ \hline
\end{tabular}
}
\newline
\caption{Comparison of recommendation performance in three real-world datasets. `*'  represents statistical significance at the 0.95 level. Improvement percentages are computed over the best baseline model (including the ablation studies) for each metric.}
\label{real-world}
\end{table*}

Note that, similarly to all other archival datasets in the field of recommender systems, we can only observe a small portion of the ratings or reward signal between all available user-item pairs. This is problematic for unbiased offline evaluations of recommendation models, especially reinforcement learning-based ones, which require the ground truth  reward for all possible user-item pairs. Therefore, in our real-world experiment, we first impute the missing reward values by fitting our reward function as shown in the simulation experiment (i.e., Eq.(\ref{eqn:R})) on the three real-world datasets. Different from the simulation experiment where we draw the values of $r_u$, $r_i$, and $e_{ij}$ from the predetermined normal distribution, these values are determined by fitting the reward function on the observed reward in these datasets. In addition, the user and item embeddings are generated from the offline datasets using the Neural Collaborative Filtering (NCF) algorithm \cite{he2017neural}, rather than generated randomly as in the simulation experiment, due to the powerfulness and popularity of NCF-based methods shown in the literature. We then report the experimental results as the average of 10 independent runs. The entire model is trained in Python using Tensorflow as the backend on an MX450 GPU.

\subsection{Experiment Results}
Table \ref{real-world} shows the performance of our proposed model as well as the baseline models on the three real-world datasets. Similar to our findings in the simulation experiment, our proposed hierarchical reinforcement learning-based model achieves significant recommendation performance improvements over all selected state-of-the-art baselines, and in terms of all four evaluation metrics (Average Reward, Hit Rate@10, Diversity, Novelty). In addition, similar to the results in Section \ref{sec:simulation_results}, our model also significantly outperforms the ablated alternatives, further indicating the importance of each component in our model design.

\section{Conclusions}
Balancing between exploiting users' known interests, and exploring to help users discover new interests, is critical to the design of modern industrial recommender systems. While extensive works have been proposed to address the challenge through classic bandits or reinforcement learning-based approaches, they do not explicitly capture user novelty-seeking intent, nor adapt the recommendation policy accordingly. We argue that hierarchical structure exists in users' novelty-seeking intent, and proposed a novel hierarchical reinforcement learning-based model to capture such user novelty-seeking intent. Our hierarchical agent includes a Session-DDPG to model user session-level novelty-seeking intent and produce session policy/goal to guide an interaction-DQN agent to make individual recommendations.

We conducted extensive simulation studies and experiments on three industrial datasets, where we observed significant recommendation performance improvements over selected state-of-the-art recommendation models. The simulation study also validates that our hierarchical RL-based model can indeed recover the hierarchical user novelty-seeking intent as constructed. We also performed extensive ablation studies to shed light on the importance of different components used in our model. We find that adopting deep hierarchical reinforcement learning methods significantly would increase the performance of recommender systems, and that it is beneficial to explicitly model the novelty-based and diversity-based metrics in the design of the reward functions used for learning. We hope the work can inspire future research on discovering even better user exploration techniques. 

For future work, we plan to strengthen our results further by studying the impact of our proposed model in different recommendation setups where the structure of user novelty-seeking intent differs, including e-commerce platforms with monetary transactions vs organic content platforms. We also plan to systematically study the long-term impact of our method. 

\bibliographystyle{ACM-Reference-Format}
\bibliography{sample-base}


\begin{thebibliography}{46}

%%% the Web address.

\ifx \showCODEN    \undefined \def \showCODEN     #1{\unskip}     \fi
\ifx \showDOI      \undefined \def \showDOI       #1{#1}\fi
\ifx \showISBNx    \undefined \def \showISBNx     #1{\unskip}     \fi
\ifx \showISBNxiii \undefined \def \showISBNxiii  #1{\unskip}     \fi
\ifx \showISSN     \undefined \def \showISSN      #1{\unskip}     \fi
\ifx \showLCCN     \undefined \def \showLCCN      #1{\unskip}     \fi
\ifx \shownote     \undefined \def \shownote      #1{#1}          \fi
\ifx \showarticletitle \undefined \def \showarticletitle #1{#1}   \fi
\ifx \showURL      \undefined \def \showURL       {\relax}        \fi
% invisible to TeX
\providecommand\bibfield[2]{#2}
\providecommand\bibinfo[2]{#2}
\providecommand\natexlab[1]{#1}
\providecommand\showeprint[2][]{arXiv:#2}

\bibitem[Adamopoulos and Tuzhilin(2014)]%
        {adamopoulos2014unexpectedness}
\bibfield{author}{\bibinfo{person}{Panagiotis Adamopoulos} {and}
  \bibinfo{person}{Alexander Tuzhilin}.} \bibinfo{year}{2014}\natexlab{}.
\newblock \showarticletitle{On unexpectedness in recommender systems: Or how to
  better expect the unexpected}.
\newblock \bibinfo{journal}{\emph{ACM Transactions on Intelligent Systems and
  Technology (TIST)}} \bibinfo{volume}{5}, \bibinfo{number}{4}
  (\bibinfo{year}{2014}), \bibinfo{pages}{1--32}.
\newblock


\bibitem[Adomavicius and Tuzhilin(2005)]%
        {adomavicius2005toward}
\bibfield{author}{\bibinfo{person}{Gediminas Adomavicius} {and}
  \bibinfo{person}{Alexander Tuzhilin}.} \bibinfo{year}{2005}\natexlab{}.
\newblock \showarticletitle{Toward the next generation of recommender systems:
  A survey of the state-of-the-art and possible extensions}.
\newblock \bibinfo{journal}{\emph{IEEE transactions on knowledge and data
  engineering}} \bibinfo{volume}{17}, \bibinfo{number}{6}
  (\bibinfo{year}{2005}), \bibinfo{pages}{734--749}.
\newblock


\bibitem[Afsar et~al\mbox{.}(2021)]%
        {afsar2021reinforcement}
\bibfield{author}{\bibinfo{person}{M~Mehdi Afsar}, \bibinfo{person}{Trafford
  Crump}, {and} \bibinfo{person}{Behrouz Far}.}
  \bibinfo{year}{2021}\natexlab{}.
\newblock \showarticletitle{Reinforcement learning based recommender systems: A
  survey}.
\newblock \bibinfo{journal}{\emph{ACM Computing Surveys (CSUR)}}
  (\bibinfo{year}{2021}).
\newblock


\bibitem[Agrawal and Goyal(2012)]%
        {agrawal2012analysis}
\bibfield{author}{\bibinfo{person}{Shipra Agrawal} {and} \bibinfo{person}{Navin
  Goyal}.} \bibinfo{year}{2012}\natexlab{}.
\newblock \showarticletitle{Analysis of thompson sampling for the multi-armed
  bandit problem}. In \bibinfo{booktitle}{\emph{Conference on learning
  theory}}. JMLR Workshop and Conference Proceedings, \bibinfo{pages}{39--1}.
\newblock


\bibitem[Antaris and Rafailidis(2021)]%
        {antaris2021sequence}
\bibfield{author}{\bibinfo{person}{Stefanos Antaris} {and}
  \bibinfo{person}{Dimitrios Rafailidis}.} \bibinfo{year}{2021}\natexlab{}.
\newblock \showarticletitle{Sequence adaptation via reinforcement learning in
  recommender systems}. In \bibinfo{booktitle}{\emph{Proceedings of the 15th
  ACM Conference on Recommender Systems}}. \bibinfo{pages}{714--718}.
\newblock


\bibitem[Auer et~al\mbox{.}(2002)]%
        {auer2002finite}
\bibfield{author}{\bibinfo{person}{Peter Auer}, \bibinfo{person}{Nicolo
  Cesa-Bianchi}, {and} \bibinfo{person}{Paul Fischer}.}
  \bibinfo{year}{2002}\natexlab{}.
\newblock \showarticletitle{Finite-time analysis of the multiarmed bandit
  problem}.
\newblock \bibinfo{journal}{\emph{Machine learning}} \bibinfo{volume}{47},
  \bibinfo{number}{2} (\bibinfo{year}{2002}), \bibinfo{pages}{235--256}.
\newblock


\bibitem[Bacon et~al\mbox{.}(2017)]%
        {bacon2017option}
\bibfield{author}{\bibinfo{person}{Pierre-Luc Bacon}, \bibinfo{person}{Jean
  Harb}, {and} \bibinfo{person}{Doina Precup}.}
  \bibinfo{year}{2017}\natexlab{}.
\newblock \showarticletitle{The option-critic architecture}. In
  \bibinfo{booktitle}{\emph{Proceedings of the AAAI Conference on Artificial
  Intelligence}}, Vol.~\bibinfo{volume}{31}.
\newblock


\bibitem[Balabanovi{\'c}(1998)]%
        {balabanovic1998exploring}
\bibfield{author}{\bibinfo{person}{Marko Balabanovi{\'c}}.}
  \bibinfo{year}{1998}\natexlab{}.
\newblock \showarticletitle{Exploring versus exploiting when learning user
  models for text recommendation}.
\newblock \bibinfo{journal}{\emph{User Modeling and User-Adapted Interaction}}
  \bibinfo{volume}{8}, \bibinfo{number}{1} (\bibinfo{year}{1998}),
  \bibinfo{pages}{71--102}.
\newblock


\bibitem[Chapelle and Li(2011)]%
        {chapelle2011empirical}
\bibfield{author}{\bibinfo{person}{Olivier Chapelle} {and}
  \bibinfo{person}{Lihong Li}.} \bibinfo{year}{2011}\natexlab{}.
\newblock \showarticletitle{An empirical evaluation of thompson sampling}.
\newblock \bibinfo{journal}{\emph{Advances in neural information processing
  systems}}  \bibinfo{volume}{24} (\bibinfo{year}{2011}).
\newblock


\bibitem[Chen et~al\mbox{.}(2019a)]%
        {chen2019top}
\bibfield{author}{\bibinfo{person}{Minmin Chen}, \bibinfo{person}{Alex Beutel},
  \bibinfo{person}{Paul Covington}, \bibinfo{person}{Sagar Jain},
  \bibinfo{person}{Francois Belletti}, {and} \bibinfo{person}{Ed~H Chi}.}
  \bibinfo{year}{2019}\natexlab{a}.
\newblock \showarticletitle{Top-k off-policy correction for a REINFORCE
  recommender system}. In \bibinfo{booktitle}{\emph{Proceedings of the Twelfth
  ACM International Conference on Web Search and Data Mining}}.
  \bibinfo{pages}{456--464}.
\newblock


\bibitem[Chen et~al\mbox{.}(2021)]%
        {chen2021values}
\bibfield{author}{\bibinfo{person}{Minmin Chen}, \bibinfo{person}{Yuyan Wang},
  \bibinfo{person}{Can Xu}, \bibinfo{person}{Ya Le}, \bibinfo{person}{Mohit
  Sharma}, \bibinfo{person}{Lee Richardson}, \bibinfo{person}{Su-Lin Wu}, {and}
  \bibinfo{person}{Ed Chi}.} \bibinfo{year}{2021}\natexlab{}.
\newblock \showarticletitle{Values of User Exploration in Recommender Systems}.
  In \bibinfo{booktitle}{\emph{Fifteenth ACM Conference on Recommender
  Systems}}. \bibinfo{pages}{85--95}.
\newblock


\bibitem[Chen et~al\mbox{.}(2022)]%
        {chen2022off}
\bibfield{author}{\bibinfo{person}{Minmin Chen}, \bibinfo{person}{Can Xu},
  \bibinfo{person}{Vince Gatto}, \bibinfo{person}{Devanshu Jain},
  \bibinfo{person}{Aviral Kumar}, {and} \bibinfo{person}{Ed Chi}.}
  \bibinfo{year}{2022}\natexlab{}.
\newblock \showarticletitle{Off-Policy Actor-critic for Recommender Systems}.
  In \bibinfo{booktitle}{\emph{Proceedings of the 16th ACM Conference on
  Recommender Systems}}. \bibinfo{pages}{338--349}.
\newblock


\bibitem[Chen et~al\mbox{.}(2019b)]%
        {chen2019generative}
\bibfield{author}{\bibinfo{person}{Xinshi Chen}, \bibinfo{person}{Shuang Li},
  \bibinfo{person}{Hui Li}, \bibinfo{person}{Shaohua Jiang},
  \bibinfo{person}{Yuan Qi}, {and} \bibinfo{person}{Le Song}.}
  \bibinfo{year}{2019}\natexlab{b}.
\newblock \showarticletitle{Generative adversarial user model for reinforcement
  learning based recommendation system}. In
  \bibinfo{booktitle}{\emph{International Conference on Machine Learning}}.
  PMLR, \bibinfo{pages}{1052--1061}.
\newblock


\bibitem[Cheng et~al\mbox{.}(2016)]%
        {cheng2016wide}
\bibfield{author}{\bibinfo{person}{Heng-Tze Cheng}, \bibinfo{person}{Levent
  Koc}, \bibinfo{person}{Jeremiah Harmsen}, \bibinfo{person}{Tal Shaked},
  \bibinfo{person}{Tushar Chandra}, \bibinfo{person}{Hrishi Aradhye},
  \bibinfo{person}{Glen Anderson}, \bibinfo{person}{Greg Corrado},
  \bibinfo{person}{Wei Chai}, \bibinfo{person}{Mustafa Ispir}, {et~al\mbox{.}}}
  \bibinfo{year}{2016}\natexlab{}.
\newblock \showarticletitle{Wide \& deep learning for recommender systems}. In
  \bibinfo{booktitle}{\emph{Proceedings of the 1st workshop on deep learning
  for recommender systems}}. \bibinfo{pages}{7--10}.
\newblock


\bibitem[Chung et~al\mbox{.}(2014)]%
        {chung2014empirical}
\bibfield{author}{\bibinfo{person}{Junyoung Chung}, \bibinfo{person}{Caglar
  Gulcehre}, \bibinfo{person}{KyungHyun Cho}, {and} \bibinfo{person}{Yoshua
  Bengio}.} \bibinfo{year}{2014}\natexlab{}.
\newblock \showarticletitle{Empirical evaluation of gated recurrent neural
  networks on sequence modeling}.
\newblock \bibinfo{journal}{\emph{arXiv preprint arXiv:1412.3555}}
  (\bibinfo{year}{2014}).
\newblock


\bibitem[Codevilla et~al\mbox{.}(2018)]%
        {codevilla2018end}
\bibfield{author}{\bibinfo{person}{Felipe Codevilla}, \bibinfo{person}{Matthias
  M{\"u}ller}, \bibinfo{person}{Antonio L{\'o}pez}, \bibinfo{person}{Vladlen
  Koltun}, {and} \bibinfo{person}{Alexey Dosovitskiy}.}
  \bibinfo{year}{2018}\natexlab{}.
\newblock \showarticletitle{End-to-end driving via conditional imitation
  learning}. In \bibinfo{booktitle}{\emph{2018 IEEE international conference on
  robotics and automation (ICRA)}}. IEEE, \bibinfo{pages}{4693--4700}.
\newblock


\bibitem[Embretson and Reise(2013)]%
        {embretson2013item}
\bibfield{author}{\bibinfo{person}{Susan~E Embretson} {and}
  \bibinfo{person}{Steven~P Reise}.} \bibinfo{year}{2013}\natexlab{}.
\newblock \bibinfo{booktitle}{\emph{Item response theory}}.
\newblock \bibinfo{publisher}{Psychology Press}.
\newblock


\bibitem[Givon(1984)]%
        {givon1984variety}
\bibfield{author}{\bibinfo{person}{Moshe Givon}.}
  \bibinfo{year}{1984}\natexlab{}.
\newblock \showarticletitle{Variety seeking through brand switching}.
\newblock \bibinfo{journal}{\emph{Marketing Science}} \bibinfo{volume}{3},
  \bibinfo{number}{1} (\bibinfo{year}{1984}), \bibinfo{pages}{1--22}.
\newblock


\bibitem[Guo et~al\mbox{.}(2017)]%
        {guo2017deepfm}
\bibfield{author}{\bibinfo{person}{Huifeng Guo}, \bibinfo{person}{Ruiming
  Tang}, \bibinfo{person}{Yunming Ye}, \bibinfo{person}{Zhenguo Li}, {and}
  \bibinfo{person}{Xiuqiang He}.} \bibinfo{year}{2017}\natexlab{}.
\newblock \showarticletitle{DeepFM: a factorization-machine based neural
  network for CTR prediction}.
\newblock \bibinfo{journal}{\emph{arXiv preprint arXiv:1703.04247}}
  (\bibinfo{year}{2017}).
\newblock


\bibitem[He et~al\mbox{.}(2017)]%
        {he2017neural}
\bibfield{author}{\bibinfo{person}{Xiangnan He}, \bibinfo{person}{Lizi Liao},
  \bibinfo{person}{Hanwang Zhang}, \bibinfo{person}{Liqiang Nie},
  \bibinfo{person}{Xia Hu}, {and} \bibinfo{person}{Tat-Seng Chua}.}
  \bibinfo{year}{2017}\natexlab{}.
\newblock \showarticletitle{Neural collaborative filtering}. In
  \bibinfo{booktitle}{\emph{Proceedings of the 26th international conference on
  world wide web}}. \bibinfo{pages}{173--182}.
\newblock


\bibitem[Kulkarni et~al\mbox{.}(2016)]%
        {kulkarni2016hierarchical}
\bibfield{author}{\bibinfo{person}{Tejas~D Kulkarni}, \bibinfo{person}{Karthik
  Narasimhan}, \bibinfo{person}{Ardavan Saeedi}, {and} \bibinfo{person}{Josh
  Tenenbaum}.} \bibinfo{year}{2016}\natexlab{}.
\newblock \showarticletitle{Hierarchical deep reinforcement learning:
  Integrating temporal abstraction and intrinsic motivation}.
\newblock \bibinfo{journal}{\emph{Advances in neural information processing
  systems}}  \bibinfo{volume}{29} (\bibinfo{year}{2016}).
\newblock


\bibitem[Li et~al\mbox{.}(2010)]%
        {li2010contextual}
\bibfield{author}{\bibinfo{person}{Lihong Li}, \bibinfo{person}{Wei Chu},
  \bibinfo{person}{John Langford}, {and} \bibinfo{person}{Robert~E Schapire}.}
  \bibinfo{year}{2010}\natexlab{}.
\newblock \showarticletitle{A contextual-bandit approach to personalized news
  article recommendation}. In \bibinfo{booktitle}{\emph{Proceedings of the 19th
  international conference on World wide web}}. \bibinfo{pages}{661--670}.
\newblock


\bibitem[Li et~al\mbox{.}(2016)]%
        {li2016collaborative}
\bibfield{author}{\bibinfo{person}{Shuai Li}, \bibinfo{person}{Alexandros
  Karatzoglou}, {and} \bibinfo{person}{Claudio Gentile}.}
  \bibinfo{year}{2016}\natexlab{}.
\newblock \showarticletitle{Collaborative filtering bandits}. In
  \bibinfo{booktitle}{\emph{Proceedings of the 39th International ACM SIGIR
  conference on Research and Development in Information Retrieval}}.
  \bibinfo{pages}{539--548}.
\newblock


\bibitem[Liebman et~al\mbox{.}(2019)]%
        {liebman2019right}
\bibfield{author}{\bibinfo{person}{Elad Liebman}, \bibinfo{person}{Maytal
  Saar-Tsechansky}, {and} \bibinfo{person}{Peter Stone}.}
  \bibinfo{year}{2019}\natexlab{}.
\newblock \showarticletitle{The right music at the right time: Adaptive
  personalized playlists based on sequence modeling.}
\newblock \bibinfo{journal}{\emph{MIS Quarterly}} \bibinfo{volume}{43},
  \bibinfo{number}{3} (\bibinfo{year}{2019}).
\newblock


\bibitem[Lillicrap et~al\mbox{.}(2015)]%
        {lillicrap2015continuous}
\bibfield{author}{\bibinfo{person}{Timothy~P Lillicrap},
  \bibinfo{person}{Jonathan~J Hunt}, \bibinfo{person}{Alexander Pritzel},
  \bibinfo{person}{Nicolas Heess}, \bibinfo{person}{Tom Erez},
  \bibinfo{person}{Yuval Tassa}, \bibinfo{person}{David Silver}, {and}
  \bibinfo{person}{Daan Wierstra}.} \bibinfo{year}{2015}\natexlab{}.
\newblock \showarticletitle{Continuous control with deep reinforcement
  learning}.
\newblock \bibinfo{journal}{\emph{arXiv preprint arXiv:1509.02971}}
  (\bibinfo{year}{2015}).
\newblock


\bibitem[Mnih et~al\mbox{.}(2015)]%
        {mnih2015human}
\bibfield{author}{\bibinfo{person}{Volodymyr Mnih}, \bibinfo{person}{Koray
  Kavukcuoglu}, \bibinfo{person}{David Silver}, \bibinfo{person}{Andrei~A
  Rusu}, \bibinfo{person}{Joel Veness}, \bibinfo{person}{Marc~G Bellemare},
  \bibinfo{person}{Alex Graves}, \bibinfo{person}{Martin Riedmiller},
  \bibinfo{person}{Andreas~K Fidjeland}, \bibinfo{person}{Georg Ostrovski},
  {et~al\mbox{.}}} \bibinfo{year}{2015}\natexlab{}.
\newblock \showarticletitle{Human-level control through deep reinforcement
  learning}.
\newblock \bibinfo{journal}{\emph{nature}} \bibinfo{volume}{518},
  \bibinfo{number}{7540} (\bibinfo{year}{2015}), \bibinfo{pages}{529--533}.
\newblock


\bibitem[Nachum et~al\mbox{.}(2018)]%
        {nachum2018data}
\bibfield{author}{\bibinfo{person}{Ofir Nachum},
  \bibinfo{person}{Shixiang~Shane Gu}, \bibinfo{person}{Honglak Lee}, {and}
  \bibinfo{person}{Sergey Levine}.} \bibinfo{year}{2018}\natexlab{}.
\newblock \showarticletitle{Data-efficient hierarchical reinforcement
  learning}.
\newblock \bibinfo{journal}{\emph{Advances in neural information processing
  systems}}  \bibinfo{volume}{31} (\bibinfo{year}{2018}).
\newblock


\bibitem[Pariser(2011)]%
        {pariser2011filter}
\bibfield{author}{\bibinfo{person}{Eli Pariser}.}
  \bibinfo{year}{2011}\natexlab{}.
\newblock \bibinfo{booktitle}{\emph{The filter bubble: How the new personalized
  web is changing what we read and how we think}}.
\newblock \bibinfo{publisher}{Penguin}.
\newblock


\bibitem[Qu et~al\mbox{.}(2016)]%
        {qu2016product}
\bibfield{author}{\bibinfo{person}{Yanru Qu}, \bibinfo{person}{Han Cai},
  \bibinfo{person}{Kan Ren}, \bibinfo{person}{Weinan Zhang},
  \bibinfo{person}{Yong Yu}, \bibinfo{person}{Ying Wen}, {and}
  \bibinfo{person}{Jun Wang}.} \bibinfo{year}{2016}\natexlab{}.
\newblock \showarticletitle{Product-based neural networks for user response
  prediction}. In \bibinfo{booktitle}{\emph{2016 IEEE 16th International
  Conference on Data Mining (ICDM)}}. IEEE, \bibinfo{pages}{1149--1154}.
\newblock


\bibitem[Shani et~al\mbox{.}(2005)]%
        {shani2005mdp}
\bibfield{author}{\bibinfo{person}{Guy Shani}, \bibinfo{person}{David
  Heckerman}, \bibinfo{person}{Ronen~I Brafman}, {and} \bibinfo{person}{Craig
  Boutilier}.} \bibinfo{year}{2005}\natexlab{}.
\newblock \showarticletitle{An MDP-based recommender system.}
\newblock \bibinfo{journal}{\emph{Journal of Machine Learning Research}}
  \bibinfo{volume}{6}, \bibinfo{number}{9} (\bibinfo{year}{2005}).
\newblock


\bibitem[Sun et~al\mbox{.}(2019)]%
        {sun2019debiasing}
\bibfield{author}{\bibinfo{person}{Wenlong Sun}, \bibinfo{person}{Sami
  Khenissi}, \bibinfo{person}{Olfa Nasraoui}, {and} \bibinfo{person}{Patrick
  Shafto}.} \bibinfo{year}{2019}\natexlab{}.
\newblock \showarticletitle{Debiasing the human-recommender system feedback
  loop in collaborative filtering}. In \bibinfo{booktitle}{\emph{Companion
  Proceedings of The 2019 World Wide Web Conference}}.
  \bibinfo{pages}{645--651}.
\newblock


\bibitem[Sutton and Barto(2018)]%
        {sutton2018reinforcement}
\bibfield{author}{\bibinfo{person}{Richard~S Sutton} {and}
  \bibinfo{person}{Andrew~G Barto}.} \bibinfo{year}{2018}\natexlab{}.
\newblock \bibinfo{booktitle}{\emph{Reinforcement learning: An introduction}}.
\newblock \bibinfo{publisher}{MIT press}.
\newblock


\bibitem[Van~der Maaten and Hinton(2008)]%
        {van2008visualizing}
\bibfield{author}{\bibinfo{person}{Laurens Van~der Maaten} {and}
  \bibinfo{person}{Geoffrey Hinton}.} \bibinfo{year}{2008}\natexlab{}.
\newblock \showarticletitle{Visualizing data using t-SNE.}
\newblock \bibinfo{journal}{\emph{Journal of machine learning research}}
  \bibinfo{volume}{9}, \bibinfo{number}{11} (\bibinfo{year}{2008}).
\newblock


\bibitem[Van~Hasselt et~al\mbox{.}(2016)]%
        {van2016deep}
\bibfield{author}{\bibinfo{person}{Hado Van~Hasselt}, \bibinfo{person}{Arthur
  Guez}, {and} \bibinfo{person}{David Silver}.}
  \bibinfo{year}{2016}\natexlab{}.
\newblock \showarticletitle{Deep reinforcement learning with double
  q-learning}. In \bibinfo{booktitle}{\emph{Proceedings of the AAAI conference
  on artificial intelligence}}, Vol.~\bibinfo{volume}{30}.
\newblock


\bibitem[Wang et~al\mbox{.}(2022)]%
        {wang2022surrogate}
\bibfield{author}{\bibinfo{person}{Yuyan Wang}, \bibinfo{person}{Mohit Sharma},
  \bibinfo{person}{Can Xu}, \bibinfo{person}{Sriraj Badam},
  \bibinfo{person}{Qian Sun}, \bibinfo{person}{Lee Richardson},
  \bibinfo{person}{Lisa Chung}, \bibinfo{person}{Ed~H Chi}, {and}
  \bibinfo{person}{Minmin Chen}.} \bibinfo{year}{2022}\natexlab{}.
\newblock \showarticletitle{Surrogate for Long-Term User Experience in
  Recommender Systems}. In \bibinfo{booktitle}{\emph{Proceedings of the 28th
  ACM SIGKDD Conference on Knowledge Discovery and Data Mining}}.
  \bibinfo{pages}{4100--4109}.
\newblock


\bibitem[Wang et~al\mbox{.}(2016)]%
        {wang2016dueling}
\bibfield{author}{\bibinfo{person}{Ziyu Wang}, \bibinfo{person}{Tom Schaul},
  \bibinfo{person}{Matteo Hessel}, \bibinfo{person}{Hado Hasselt},
  \bibinfo{person}{Marc Lanctot}, {and} \bibinfo{person}{Nando Freitas}.}
  \bibinfo{year}{2016}\natexlab{}.
\newblock \showarticletitle{Dueling network architectures for deep
  reinforcement learning}. In \bibinfo{booktitle}{\emph{International
  conference on machine learning}}. PMLR, \bibinfo{pages}{1995--2003}.
\newblock


\bibitem[Watkins and Dayan(1992)]%
        {watkins1992q}
\bibfield{author}{\bibinfo{person}{Christopher~JCH Watkins} {and}
  \bibinfo{person}{Peter Dayan}.} \bibinfo{year}{1992}\natexlab{}.
\newblock \showarticletitle{Q-learning}.
\newblock \bibinfo{journal}{\emph{Machine learning}} \bibinfo{volume}{8},
  \bibinfo{number}{3} (\bibinfo{year}{1992}), \bibinfo{pages}{279--292}.
\newblock


\bibitem[Xiao and Benbasat(2007)]%
        {xiao2007commerce}
\bibfield{author}{\bibinfo{person}{Bo Xiao} {and} \bibinfo{person}{Izak
  Benbasat}.} \bibinfo{year}{2007}\natexlab{}.
\newblock \showarticletitle{E-commerce product recommendation agents: Use,
  characteristics, and impact}.
\newblock \bibinfo{journal}{\emph{MIS quarterly}} (\bibinfo{year}{2007}),
  \bibinfo{pages}{137--209}.
\newblock


\bibitem[Xiao and Wang(2021)]%
        {xiao2021general}
\bibfield{author}{\bibinfo{person}{Teng Xiao} {and} \bibinfo{person}{Donglin
  Wang}.} \bibinfo{year}{2021}\natexlab{}.
\newblock \showarticletitle{A general offline reinforcement learning framework
  for interactive recommendation}. In \bibinfo{booktitle}{\emph{Proceedings of
  the AAAI Conference on Artificial Intelligence}}, Vol.~\bibinfo{volume}{35}.
  \bibinfo{pages}{4512--4520}.
\newblock


\bibitem[Xie et~al\mbox{.}(2021)]%
        {xie2021hierarchical}
\bibfield{author}{\bibinfo{person}{Ruobing Xie}, \bibinfo{person}{Shaoliang
  Zhang}, \bibinfo{person}{Rui Wang}, \bibinfo{person}{Feng Xia}, {and}
  \bibinfo{person}{Leyu Lin}.} \bibinfo{year}{2021}\natexlab{}.
\newblock \showarticletitle{Hierarchical reinforcement learning for integrated
  recommendation}. In \bibinfo{booktitle}{\emph{Proceedings of the AAAI
  Conference on Artificial Intelligence}}, Vol.~\bibinfo{volume}{35}.
  \bibinfo{pages}{4521--4528}.
\newblock


\bibitem[Xin et~al\mbox{.}(2020)]%
        {xin2020self}
\bibfield{author}{\bibinfo{person}{Xin Xin}, \bibinfo{person}{Alexandros
  Karatzoglou}, \bibinfo{person}{Ioannis Arapakis}, {and}
  \bibinfo{person}{Joemon~M Jose}.} \bibinfo{year}{2020}\natexlab{}.
\newblock \showarticletitle{Self-supervised reinforcement learning for
  recommender systems}. In \bibinfo{booktitle}{\emph{Proceedings of the 43rd
  International ACM SIGIR conference on research and development in Information
  Retrieval}}. \bibinfo{pages}{931--940}.
\newblock


\bibitem[Zhang et~al\mbox{.}(2022)]%
        {zhang2022multi}
\bibfield{author}{\bibinfo{person}{Qihua Zhang}, \bibinfo{person}{Junning Liu},
  \bibinfo{person}{Yuzhuo Dai}, \bibinfo{person}{Yiyan Qi},
  \bibinfo{person}{Yifan Yuan}, \bibinfo{person}{Kunlun Zheng},
  \bibinfo{person}{Fan Huang}, {and} \bibinfo{person}{Xianfeng Tan}.}
  \bibinfo{year}{2022}\natexlab{}.
\newblock \showarticletitle{Multi-Task Fusion via Reinforcement Learning for
  Long-Term User Satisfaction in Recommender Systems}. In
  \bibinfo{booktitle}{\emph{Proceedings of the 28th ACM SIGKDD Conference on
  Knowledge Discovery and Data Mining}}. \bibinfo{pages}{4510--4520}.
\newblock


\bibitem[Zhao et~al\mbox{.}(2019)]%
        {zhao2019deep}
\bibfield{author}{\bibinfo{person}{Xiangyu Zhao}, \bibinfo{person}{Long Xia},
  \bibinfo{person}{Jiliang Tang}, {and} \bibinfo{person}{Dawei Yin}.}
  \bibinfo{year}{2019}\natexlab{}.
\newblock \showarticletitle{Deep reinforcement learning for search,
  recommendation, and online advertising: a survey}.
\newblock \bibinfo{journal}{\emph{ACM sigweb newsletter}}
  \bibinfo{number}{Spring} (\bibinfo{year}{2019}), \bibinfo{pages}{1--15}.
\newblock


\bibitem[Zheng et~al\mbox{.}(2018)]%
        {zheng2018drn}
\bibfield{author}{\bibinfo{person}{Guanjie Zheng}, \bibinfo{person}{Fuzheng
  Zhang}, \bibinfo{person}{Zihan Zheng}, \bibinfo{person}{Yang Xiang},
  \bibinfo{person}{Nicholas~Jing Yuan}, \bibinfo{person}{Xing Xie}, {and}
  \bibinfo{person}{Zhenhui Li}.} \bibinfo{year}{2018}\natexlab{}.
\newblock \showarticletitle{DRN: A deep reinforcement learning framework for
  news recommendation}. In \bibinfo{booktitle}{\emph{Proceedings of the 2018
  world wide web conference}}. \bibinfo{pages}{167--176}.
\newblock


\bibitem[Zhou et~al\mbox{.}(2018)]%
        {zhou2018deep}
\bibfield{author}{\bibinfo{person}{Guorui Zhou}, \bibinfo{person}{Xiaoqiang
  Zhu}, \bibinfo{person}{Chenru Song}, \bibinfo{person}{Ying Fan},
  \bibinfo{person}{Han Zhu}, \bibinfo{person}{Xiao Ma},
  \bibinfo{person}{Yanghui Yan}, \bibinfo{person}{Junqi Jin},
  \bibinfo{person}{Han Li}, {and} \bibinfo{person}{Kun Gai}.}
  \bibinfo{year}{2018}\natexlab{}.
\newblock \showarticletitle{Deep interest network for click-through rate
  prediction}. In \bibinfo{booktitle}{\emph{Proceedings of the 24th ACM SIGKDD
  international conference on knowledge discovery \& data mining}}.
  \bibinfo{pages}{1059--1068}.
\newblock


\bibitem[Zou et~al\mbox{.}(2019)]%
        {zou2019reinforcement}
\bibfield{author}{\bibinfo{person}{Lixin Zou}, \bibinfo{person}{Long Xia},
  \bibinfo{person}{Zhuoye Ding}, \bibinfo{person}{Jiaxing Song},
  \bibinfo{person}{Weidong Liu}, {and} \bibinfo{person}{Dawei Yin}.}
  \bibinfo{year}{2019}\natexlab{}.
\newblock \showarticletitle{Reinforcement learning to optimize long-term user
  engagement in recommender systems}. In \bibinfo{booktitle}{\emph{Proceedings
  of the 25th ACM SIGKDD International Conference on Knowledge Discovery \&
  Data Mining}}. \bibinfo{pages}{2810--2818}.
\newblock


\end{thebibliography}

\end{document}